# A very low-current electromagnetic induction experiment enhanced by acoustic means

*Santiago Ortuño-Molina[1], Rod Milbrandt[2], Juan C. Castro-Palacio[1], Juan A. Monsoriu[1]*

[1]Centro de Tecnologías Físicas, Universitat Politècnica de València, Camino de Vera, s/n, 46022 València, Spain

[2]Faculty of Physics and Engineering, Rochester Community and Technical College, Rochester, Minnesota, The United States.

In July 2024, two of the authors of this article were attending the 2024 Summer Meeting of the American Association of Physics Teachers (AAPT) in Boston. On this occasion, we had the opportunity to attend a very interesting physics demonstration show at the Harvard University Sanders Theatre. One of the experiments involved transmitting music through two large coils, one connected to a music player and another to a speaker. The coils faced each other and were separated by a short distance. The audio signal was transmitted from one coil to the other via electromagnetic induction and as a result the music could be heard through the speaker on the opposite side.

Upon returning home, we considered the possibility of carrying out this experiment using mobile phones and small coils, making it portable and engaging for students. This work aligns with over a decade of previous research on the use of smartphone sensors in physics teaching laboratory[1-5].

After some discussion we came up with a design (Figure 1) that includes smartphones used to generate sinusoidal signals, copper coils of 500 turns each, and small computer speakers. Another smartphone was used to collect the data. The amplification system of the speakers is key to the success of this experiment as the induced currents produced are very low (a few mA) with ordinary teaching lab equipment. This is why for induction experiments primary currents above 1-2 A are usually needed.

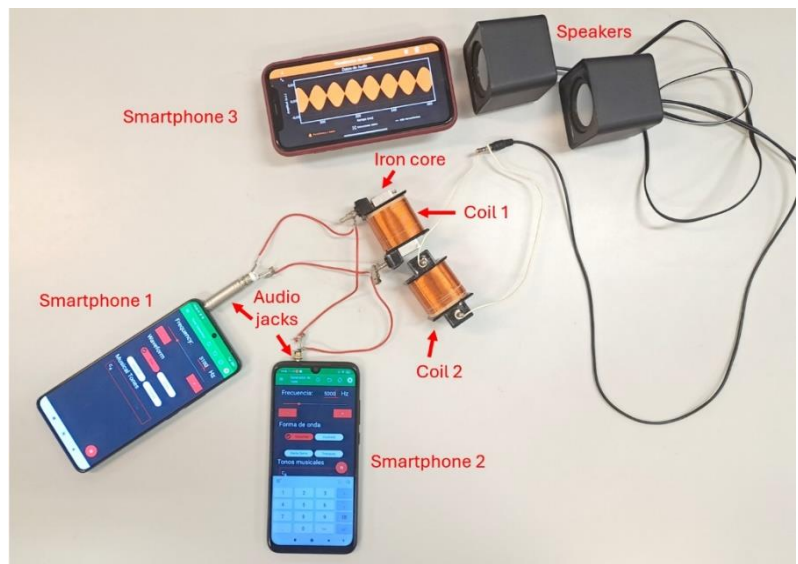

**Figure 1.** Experimental setup used in this work. Smartphones 1 and 2 are used as signal generators, whereas smartphone 3 is used as an oscilloscope.

A smartphone can be used as a signal generator by using the Tone Generator function included in the free smartphone app Physics Sensors Toolbox Suite[6]. This option can be found on the left panel of the main screen of the app. The frequency (ranging from 0 to 20 000 Hz) and the type of signal (sinusoidal, square, saw-tooth, triangular) can be chosen. The amplitude of the generated wave can be controlled by changing the volume of the phone. An audio jack has been used to connect the phone to the coil.

Using the above setup (Figure 1), we have developed a lab activity for secondary school and introductory university physics courses which consists of measuring the acoustic beats that result from superimposing two sinusoidal signals at the coil on the transmission side, and then measuring the beats from the electromagnetic induction produced at a second coil connected to a third smartphone. To collect and export the data, we used the function Audio Scope included in the free smartphone app Phyphox[7]. The authors have used a smartphone as a function generator in previous work[8] studying RLC circuits. The different manifestations of beats have been studied in other previous work, for instance, the acoustic beats between sound waves have been studied with tuning forks[9-10], speakers and microphones[11], and more recently using smartphones[12-13].

The beat phenomenon is produced when two signals of slightly different frequencies are superimposed. In the case of acoustic beats, a periodic variation of the sound intensity can be heard. Mathematically, it can be expressed as follows,

$$A_1\sin(2\pi f_1 t + \varphi_1) + A_2\sin(2\pi f_2 t + \varphi_2) = A_E(t)\sin\left(2\pi \frac{f_1+f_2}{2} + \varphi\right), \qquad (1)$$

where

$$A_E(t) = \sqrt{A_1^2 + A_2^2 + 2A_1 A_2 \cos(2\pi|f_1 - f_2|t + \varphi_E)}. \qquad (2)$$

It can be seen that the frequency of the envelope, $A_E(t)$, which is the frequency of the beats, is just $f_{\text{beat}} = |f_1 - f_2|$.

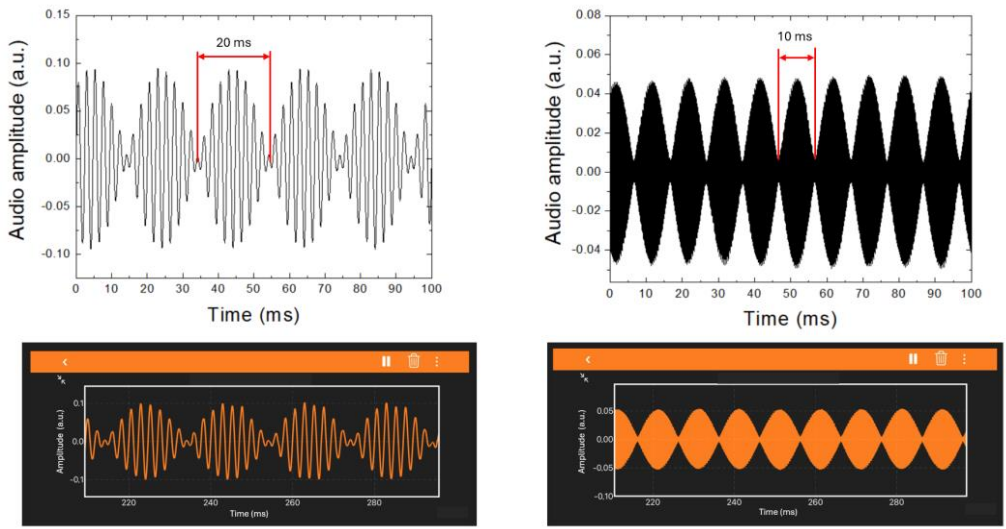

**Figure 2.** Acoustic beats resulting from sinusoidal signals of 400 and 450 Hz frequency (panels on the left) and from signals of 5000 and 5100 Hz frequency (panels on the right). The graphs from the exported experimental data (upper row) and screenshots of the smartphones' screens (lower row) are included.

Figure 2 shows the observed beats for two pairs of frequencies, 400 and 450 Hz (panels on the lefthand side), and 5000 and 5100 Hz (panels on the righthand side). The upper figure of each panel includes the plots obtained from the saved data file. The pictures below the graphs are smartphone screenshots (using the Audio Scope option on the phyphox app[6]). For the first pair, using 400 and 450 Hz, the beat frequency is $f_{\text{beat}} = 50$ Hz and for the second pair, using 5000 and 5100 Hz, $f_{\text{beat}} = 100$ Hz. The graphs in Figure 2 (upper row) show the period corresponding to each beat frequency ($T_{\text{beat}} = 1/f_{\text{beat}}$), 20 ms and 10 ms, respectively.

According to Faraday's Law of Induction[14], a time variation of the flux of the magnetic field vector across a surface gives rise to an electromotive force. Let us consider the case of a constant magnetic field over the area for a given moment of time,

$$\varepsilon_{ind} = -\frac{d}{dt}\Phi_{\vec{B}} = -\frac{d}{dt}\vec{B}\cdot\vec{S} = -\frac{d}{dt}|\vec{B}||\vec{S}|\cos(\vec{B},\vec{S}), \tag{3}$$

where $\vec{B}$ is the magnetic field vector and $\vec{S}$ is the area vector whose direction is perpendicular to the planar area $S$, considered constant for this case.

The superposition of sinusoidal signals (beats) in the experiments described above is transmitted via electromagnetic induction between the coils. A variable magnetic field is produced around the coil which is connected to the smartphones, due to the variable electric current circulating around the copper wire. The magnetic field in a coil is mainly along its longitudinal axis. According to Faraday's Law as stated in Eq. 3, a variable magnetic field across the area of the secondary coil, which is connected to the loudspeakers, will induce currents which oscillate with the same frequency as in the primary coil.

Based on equation 1, we can also ask the students to explore different situations qualitatively. For this purpose, the setup in figure 1 can be used. For instance, keeping the area constant, we can explore the case of changing the angle between $\vec{B}$ (produced at the first coil) and the area vector $\vec{S}$ (of the second coil). Starting from both coils with the axes aligned, one of them can be turned to reduce the effective area of the flux. This change would decrease the flux. The highest sound intensity from the speakers will be obtained when the coil axes are aligned. However, if the axes of the coils are perpendicular to each other, almost no sound can be heard from the speakers as the magnetic field lines produced at the first coil are parallel to the cross section of the second coil and therefore the resulting magnetic flux is zero. Another situation to explore is what happens when the coils are moved away from each other. The magnetic field decreases inversely with the separation distance. Then, on the screen of a third smartphone (Figure 2), we see that the sound amplitude decreases with increasing separation between the coils. A sound level decrease can be heard at the loudspeakers, although it is known that it is not linearly proportional to the sound intensity.

An alternative experiment can be set up for the case when only a Bluetooth connection is available on the smartphone. A similar experimental setup was used with two concentric coils of length 12 cm (PASCO model SE-8653) and a steel core (Figure 3). Audio signal from an iPhone was sent via a Bluetooth receiver (1Mii model B06) to the outer coil and the inner coil was connected to two computer speakers. In this case, when the coil overlap is changed the sound volume changes audibly and can again be analyzed using the Audioscope phyphox app[6].

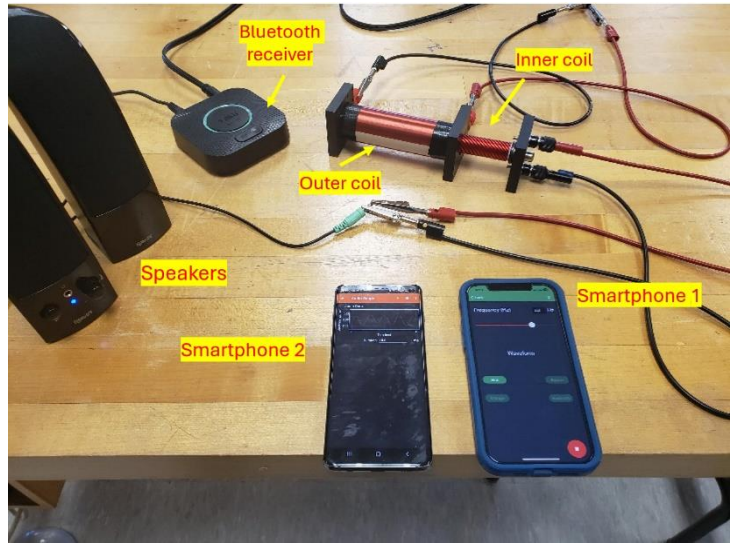

**Figure 3.** Bluetooth transmission of audio signal to concentric coils with adjustable overlap. In the figure, smartphone 1 is used to generate the audio signal which is sent via a Bluetooth receiver to the outer coil. An inner coil is connected to two computer speakers. The sound produced by the speakers is displayed by smartphone 2.

To wrap up the lab session with a fun experiment, we ask the students to use the first setup (Figure 1) to play music which can be surprising and fun for them. Figure 4a shows a smartphone on which the song "We Will Rock You" by the band Queen is played. This smartphone is connected to a coil. Then, on the transmission side, there are speakers connected to another coil. The music can be clearly heard from the speakers[15].

An even more surprising effect is shown in Figure 4 (panels c and d), where one of the earphone cables is wound to form a small coil and then used to listen to the music by holding it close to the coil. We use earphones that include a microphone. A picture of these earphones and a schematic representation of their circuit are included in Figure 5. For the experiment to work, we need to close the circuit. The optimal case is when the left and right audio ends are connected as the sound can then be heard from both speakers. The current path is depicted for this case in Figure 5a. If the right audio and the ground are connected only the right speaker will emit sound. However, when the microphone is connected to the ground or the left audio is connected to the ground, no sound is heard from the speakers. If the wires between the speakers and the microphone are looped to make coils, no sound will be produced either as antiparallel currents will originate with almost no net current resulting.

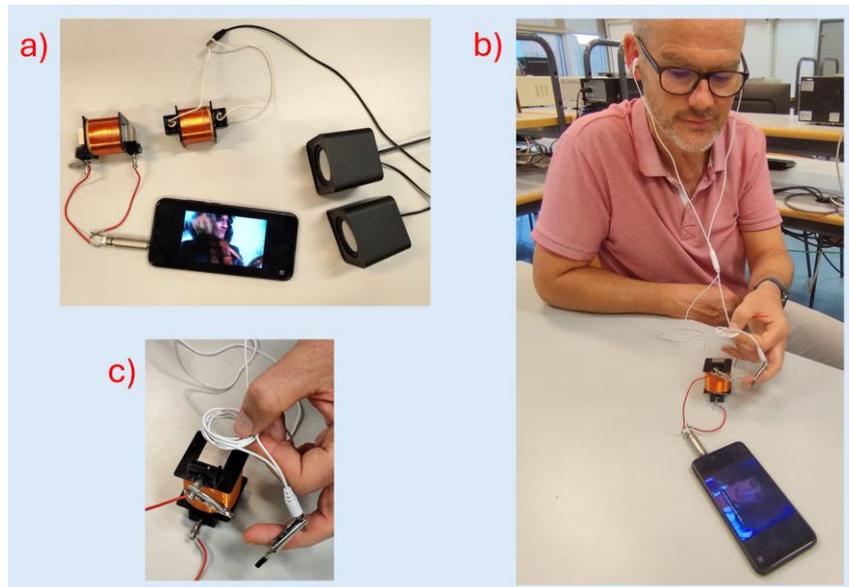

**Figure 4.** Transmission of music using speakers (Figure 4a) and earphones (Figure 4b and c).

Figure 5b shows a generic circuit diagram for the common case of earphones without a microphone. In this case, no sound is expected to be heard from the speakers as again, antiparallel currents will be produced at the wound cables. In this case, one could use another coil and clip to the headphone jack as shown in Figure 4a.

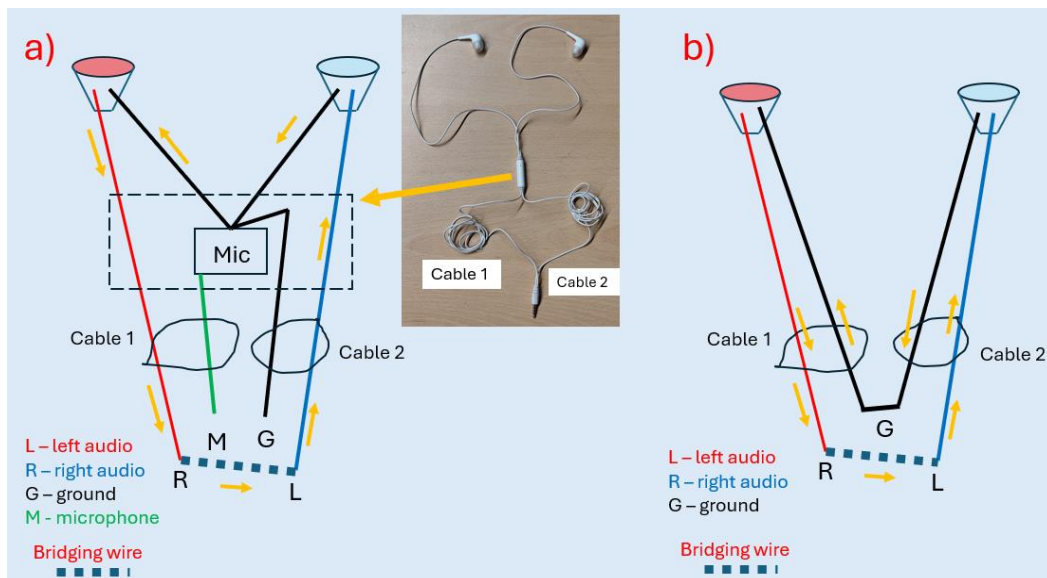

**Figure 5.** Photo and schematic representation of the earphones used in this work (panel a) which include a microphone. Panel b) includes a generic circuit diagram for earphones without a microphone.

**Some pedagogical considerations**

The proposed experiments combine several physics concepts which are studied in secondary schools such as magnetic flux, electromagnetic induction and acoustic waves. Using smartphones, which are very familiar devices for students, makes the experiments more portable and opens the possibility of carrying them out at home since the coils can be easily made. The same setup can be used in introductory college physics courses to study electromagnetic induction with more detail or RL circuits where the resistance (impedance) depends on the frequency.

These experiments aim to get the students closer to electromagnetic induction phenomena. The setup is quite simple and is composed of materials and devices which are very familiar to the students. We know from our experience that students are more motivated when the experiments involve smartphones as measurement instruments. Results indicate that the physics concepts are better understood.

In this work, we have created a simple scenario allowing the students and teachers to experiment with and discuss induction in a safe way, using low currents and a very portable setup. The simplicity and portability make these activities well suited for experiments outside of the classroom, which is particularly welcome when special situations occur, such as distance learning during the COVID-19 pandemic.

Thus far, we have used the full version of these activities with secondary school students and in first-year college physics courses as class demonstrations. We are now exploring how best to modify the experiments to provide a more comprehensive lab activity for engineering students. This new version may include the quantitative model for the magnetic field of a coil when it is fed with alternating current using the smartphone as signal generator. In this case the full expression for the induced emf in the second coil could be stated theoretically and proven experimentally.

We have no doubt that simple, low cost, portable experiments using engaging elements such as smartphones, sound, and music, will be attractive for secondary school and first-year university physics courses.

**Acknowledgments**

The authors would like to thank the Instituto de Ciencias de la Educación (Institute of Education Sciences) at the Universitat Politècnica de València (UPV) for its support of the teaching innovation group MSEL.

**References**

1. J.C. Castro-Palacio, L. Velazquez-Abad, M.H. Gimenez and J.A. Monsoriu, "Using a mobile, phone acceleration sensor in physics experiments on free and damped harmonic oscillations," *Am. J. Phys.* **81**, 472-475 (2013). http://hdl.handle.net/10251/54285
2. M.H. Giménez, I. Salinas, J.A. Monsoriu and Juan C. Castro-Palacio, "Direct Visualization of Mechanical Beats by Means of an Oscillating Smartphone," *Phys. Teach.* **55**, 424-425 (2017). https://doi.org/10.1119/1.5003745

3. I. Salinas, M.H. Giménez, J.A. Monsoriu y J.A. Sans, "Demostration of the parallel axis theorem through a smartphone," *Phys. Teach.* **57**, 340 (2019). https://doi.org/10.1119/1.5098929
4. I. Salinas, M. Monteiro, A. Martí and J.A. Monsoriu, "Analyzing the Dynamics of a Yo-Yo Using a Smartphone Gyroscope Sensor," *Phys. Teach.* **58**, 569 (2020). https://doi.org/10.1119/10.0002379
5. C.F. Marín-Sepúlveda, J.C. Castro-Palacio, Isabel Salinas y J.A. Monsoriu, "Acoustic characterization of magnetic braking with a smartphone," *Phys. Teach.* **60**, 547-548 (2022). https://doi.org/10.1119/5.0097792
6. S. Staacks, S. Hütz, H. Heinke and C. Stampfer, "Advanced tools for smartphone-based experiments: phyphox," *Phys. Educ.* **53**, 045009 (2018). https://doi.org/10.1088/1361-6552/aac05e
7. V. Sotware 2020, "Physics toolbox," downloaded from Google Play on 1-7-2023 (available at: https://www.vieyrasoftware.net/)
8. I. Torriente-García, A. C Martí, M. Monteiro, C. Stari, J. C Castro-Palacio and J. A Monsoriu, "RLC series circuit made simple and portable with smartphones," *Phys. Educ.* **59**, 015016 (2024). https://doi.org/10.1088/1361-6552/ad04fb
9. P. Gluck, "You can illustrate beats without the usual tuning forks," *Phys. Educ.* **39**, 241 (2004). https://doi.org/10.1088/0031-9120/39/3/F09
10. T. B. Greenslade, "Beats produced by a moving tuning fork," *Phys. Teach.* **31**, 443 (1993). https://doi.org/10.1119/1.2343837
11. A. Ganci and S. Ganci, "The simplest demonstration on acoustic beats," *Phys. Teach.* **53**, 32 (2015). https://doi.org/10.1119/1.4904239
12. J. Kuhn, P. Vogt and M. Hirth, "Analyzing the acoustic beat with mobile devices," *Phys. Teach*. **52**, 248 (2014). https://doi.org/10.1119/1.4868948
13. M. Osorio, C. J. Pereyra, D. L. Gau and A. Laguarda, "Measuring and characterizing beat phenomena with a smartphone," *Eur. J. Phys.* **39,** 025708 (2018). https://doi.org/10.1088/1361-6404/aa9034
14. R. Resnick, D. Halliday and K.S. Krane, Physics 4th edition (Mexico, DF: CECSA, 1999).
15. Readers can view the video at TPT Online at https://tpt.peerx-press.org/ms_files/tpt/2024/12/15/01834438/01/1834438_1_data_set_24903810_sxk12x.mp4, in the "Supplementary Material" section.